\documentclass[12pt]{article}

\pdfoutput=1 

\usepackage[T1]{fontenc} 

\usepackage{CJK}
\usepackage{mathtools}
\usepackage{tikz}
\usepackage{amssymb}
\usepackage{epsfig}
\usepackage{bbm}
\usepackage{bbold}
\usepackage{tabularx,booktabs}
\usepackage{graphicx}
\usepackage{hyperref}
\usepackage{romannum}
\usepackage{graphics,graphpap}
\usepackage{graphicx}
\usepackage{amsmath,exscale,amsthm,bm}
\usepackage{yfonts}
\usepackage{mathrsfs}
\usepackage[toc]{appendix}
\usetikzlibrary{trees}
\usetikzlibrary{positioning,arrows}
\usetikzlibrary{decorations.pathmorphing}
\usetikzlibrary{decorations.markings}
\usetikzlibrary{decorations.text}
\usetikzlibrary{fit,chains,calc,shapes.geometric,intersections}
\usepackage{natbib}
\setcitestyle{square,comma,numbers,sort&compress}
\usepackage{circledsteps}

\begin{document}

\def\mc#1{\mathcal#1}
\def\a{\alpha}
\def\b{\beta}
\def\c{\chi}
\def\d{\delta}
\def\e{\epsilon}
\def\f{\phi}
\def\g{\gamma}
\def\h{\eta}
\def\i{\iota}
\def\j{\psi}
\def\k{\kappa}
\def\la{\lambda}
\def\m{\mu}
\def\n{\nu}
\def\o{\omega}
\def\p{\pi}
\def\q{\theta}
\def\r{\rho}
\def\s{\sigma}
\def\t{\tau}
\def\u{\upsilon}
\def\x{\xi}
\def\z{\zeta}
\def\D{\Delta}
\def\F{\Phi}
\def\G{\Gamma}
\def\J{\Psi}
\def\L{\Lambda}
\def\O{\Omega}
\def\P{\Pi}
\def\Q{\Theta}
\def\S{\Sigma}
\def\U{\Upsilon}
\def\X{\Xi}

\def\ve{\varepsilon}
\def\vf{\varphi}
\def\vr{\varrho}
\def\vs{\varsigma}
\def\vq{\vartheta}

\newcommand{\vev}[1]{\langle #1 \rangle}
\def\dg{\dagger}                                     
\def\ddg{\ddagger}                                   
\def\wt#1{\widetilde{#1}}                    
\def\mt{\widetilde{m}_1}
\def\mti{\widetilde{m}_i}
\def\mtj{\widetilde{m}_j}
\def\rt{\widetilde{r}_1}
\def\mtt{\widetilde{m}_2}
\def\mttt{\widetilde{m}_3}
\def\rtt{\widetilde{r}_2}
\def\mb{\overline{m}}
\def\VEV#1{\left\langle #1\right\rangle}        
\def\be{\begin{equation}}
\def\ee{\end{equation}}
\def\ds{\displaystyle}
\def\ra{\rightarrow}

\def\bea{\begin{eqnarray}}
\def\eea{\end{eqnarray}}
\def\NO{\nonumber}
\def\Bar#1{\overline{#1}}
\def\ylz{\textcolor{red}}

\def\ket#1{|\, #1 \, \rangle}    
\def\bra#1{\langle \, #1 \, |}
\def\brac#1{\langle \, #1 \,}

\title{
\vspace*{1mm}
{\bf Non-adiabatic transitions \\ in the density matrix formalism}

\author{
{\Large Pasquale Di Bari$^a$,  Shreya Pandit$^a$ and Ye-Ling Zhou$^b$}
\\
$^a$ {\it\small Physics and Astronomy,  University of Southampton,  Southampton, SO17 1BJ, U.K.} 
\\
$^b$ {\it\small School of Fundamental Physics and Mathematical Sciences}, \\ 
{\it\small Hangzhou Institute for Advanced Study, UCAS, Hangzhou 310024, China} 
}}
\maketitle \thispagestyle{empty}
\pagenumbering{arabic}

\begin{abstract}
 We show that a density matrix formalism provides a useful description of non-adiabatic transitions in two-state quantum systems.  
 Compared to a traditional Hamiltonian formalism, even in the absence of decoherence when there is full equivalence between the two, 
the density matrix formalism provides a convenient change of variables that yields a powerful general analytical solution. 
This solution nicely describes a transition regime between the well known  Landau-Zener-Stuckelberg-Majorana (LZSM) 
approximation and the extremely non-adiabatic limit.  We also extend it including decoherence in the case of open quantum systems. 
Our results have very general applications, within a large variety of problems in quantum physics, neutrino physics, cosmology.  
\end{abstract}

\newpage

\section{Introduction}

Non-adiabatic transitions between quantum states in systems with a time-dependent Hamiltonian have many important applications. 
Historically, the problem has been first considered in atomic collisions and molecular dynamics by Landau
\cite{Landau:1932wdt}, Zener \cite{Zener:1932ws}, St\"{u}ckelberg \cite{Stuck1932} and Majorana \cite{Majorana:1932ga} 
and, therefore, they are usually referred to as LZSM transitions.  
More recently, they have found application in the description of many other processes in quantum physics such as 
nuclear magnetic resonance \cite{boucher}, quantum chemistry \cite{suchan}, quantum computing \cite{Shevchenko:2010ms,Blais:2020wjs}.  

In particle physics, neutrino mixing  leads to neutrino oscillations that have been observed in a variety of different physical environments: 
solar, atmospheric and baseline experiments. A description of medium effects,  clearly observed in all these three environments,
requires a time-dependent Hamiltonian operator \cite{Wolfenstein:1977ue,Mikheyev:1985zog}. In these three cases an adiabatic regime 
provides a good description, mainly because the three neutrino mixing angles  are sufficiently large that non-adiabatic effects can be neglected. 
However,  active-sterile neutrino mixing  could be still present, and  to be investigated, and give sub-dominant effects, for example in
solar neutrinos \cite{Bahcall:2002ij} and in this case non-adiabatic effects might become important.
Neutrinos  are also produced in supernovae,  very powerful explosions occurring at the end of the life of some stars  \cite{Raffelt:1999tx}.
Neutrinos of all flavours are produced  through various processes, for example from electron-positron annihilations. 
In 1987 the Kamiokande-II  \cite{Kamiokande-II:1987idp} and the IMB \cite{Bionta:1987qt} detectors observed a few electron neutrinos from SN1987A.  
Although only about 24 events of the estimated $10^{28}$ neutrinos that were generated by the SN explosion and 
passed through the Earth were detected, these few events already provided precious information about the physics of 
core-collapse SNe. 
A future galactic supernova explosion
would provide a chance to observe a much higher number of neutrinos of all flavours. The flavour spectrum, at different energies,
would be very sensitive to neutrino mixing parameters, matter effects, properties of the SN explosion \cite{Dighe:1999bi}. 
In this  case it is not excluded that non-adiabatic transitions might play a role.  Medium effects will also play an important role
in long baseline experiments such as DUNE and also in this case non-adiabatic transitions might give some detectable effect.

There are also various  interesting (yet to be tested) non-standard particle mixing processes that might be important in the early universe and 
where non-adiabatic transitions play an important role: 
\begin{itemize}
\item[(i)] Active-sterile neutrino mixing with a small mixing angle could produce either dark radiation 
\cite{Barbieri:1989ti,Enqvist:1991qj,Foot:1995qk,DiBari:1999ha} or dark matter \cite{Dodelson:1993je}. It could  
also alter the cosmic neutrino background properties \cite{DiBari:2001jk} and,  
recently, it has been noticed that it could also provide an explanation for the excess radio background mystery \cite{Dev:2025ufo}.  
\item[(ii)] Sterile-sterile neutrino mixing has been proposed as a production mechanism 
for very heavy dark matter \cite{Anisimov:2008gg} and, in combination with a primordial strong first order 
phase transition, also of gravitational waves \cite{DiBari:2020bvn}. 
 \item[(iii)] Cosmological effects of kinetic mixing between a photon and a dark photon has been intensively studied in the last years \cite{Mirizzi:2009iz}, especially in relation to
 dark photon as a dark matter candidate.
 \end{itemize}

A density matrix formalism is usually introduced to  describe how a system interacts with the environment (open quantum system)
accounting for quantum decoherence \cite{Lindblad:1975ef,Gorini:1975nb,franke}.   For example, in active-sterile neutrino oscillations in the early universe it 
provides the right formalism to describe a correct transition between a collisional regime and a regime driven by oscillations \cite{Bell:1998ds}.
However, at the same time it has also been found convenient to describe a non-adiabatic regime even in the absence of 
decoherence induced by collisions. For example, this is the case both in active-sterile neutrino mixing
in the early universe \cite{DiBari:2001jk} and in very heavy dark matter production from sterile-sterile neutrino mixing \cite{DiBari:2019zcc}.

In this paper we generalise this use of the density matrix formalism. Although we consider systems described by 
two-state quantum states, our discussion can be generalised to quantum systems living in a higher dimensional Hilbert space. 
Our main goal is to derive analytical insight  in the description of non-adiabatic transitions occurring during 
the time-evolution of quantum systems, since there seems to be a growing interest in this problem within different fields, as discussed. 
In this respect we derive an analytical expression describing such transitions generalising the traditional 
LZSM approximation, also accounting for decoherence in the case of open quantum systems. 

The paper is structured in the following way. 
In Section 2 we review a description of a two-state quantum system within the density matrix formalism.
This is useful to set the notation. 
In Section 3 we discuss how the Hamiltonian operator can be conveniently decomposed  into a time-independent and a time-dependent part, deriving
an expression for the (time-dependent) mixing angle. 
In Section 4 we discuss, for illustrative purposes, the simplest case of a time-independent Hamiltonian. 
In Section 5 we consider time-dependence in the adiabatic regime. This can be regarded as the next-to-simplest case.
In Section 6 we discuss transitions in the  non-adiabatic regime deriving an expression for the conversion probability.
In Section 7 we include decoherence and generalise the expression for the conversion probability.
Finally, in Section 8 we draw our conclusions.  In the Appendix we review the derivation of the LZSM approximation
in the Hamiltonian formalism (the original derivation by Zener 
and a general approach based on the semi-classical approximation equivalent to the derivation by Landau). 


\section{Density matrix formalism}

Let us first briefly review how the time evolution of a two-quantum system is described within the density matrix formalism. 
The Schr\"{o}dinger equation for the time evolution of the quantum state,
\be\label{Scheq}
i \, {\partial \over \partial t}\ket{\psi(t)} = \hat{H}(t) \ket{\psi(t)} \,  ,
\ee
is fully equivalent to the von Neumann equation for the (time-dependent) density matrix operator 
$\hat{\rho}(t) \equiv \ket{\psi(t)}\bra{\psi(t)}$, 
\be\label{vonneumann}
{\partial \hat\rho(t) \over \partial t} = - i \, \left[\hat H(t),\hat\rho(t) \right] \,  ,
\ee
the analogue of the Liouville equation in classical mechanics. In general, the Hamiltonian operator is also 
explicitly depending on time. The density operator is represented, 
in some generic basis $\{ \ket{\lambda_i} \}_{i=1,\dots,n}$, by the {\em density matrix} $\rho(t)$
with matrix elements $\rho_{ij}(t) \equiv \bra{\lambda_i} \hat \rho(t) \ket{\lambda_j}$ 
($i,j=1\,\dots,n$).\footnote{Normalised states ($\brac{\psi}\ket{\psi}= 1$)
imply ${\rm Tr}[\rho] = 1$, as expected, since $\rho_{ii} = |\brac{\lambda_i}\ket{\psi}|^2$ gives the probability to find the state 
$|\psi \rangle$ in the eigenstate $\ket{\lambda_i}$. In addition, for a pure state, in the absence of decoherence, $\rho$ is idempotent,
i.e., $\rho^2 = \rho$.}
In this case, we can write the von Neumann equation for the density matrix as
\be\label{vonneumann2}
{\partial \rho(t) \over \partial t} = - i \, \left[H(t) , \rho(t) \right] \,  ,
\ee
where $H(t)$ is the Hamiltonian matrix with elements $H_{ij}(t) \equiv \bra{\lambda_i} \hat H(t) \ket{\lambda_j}$.
As anticipated in the introduction, we focus, for simplicity, on the simplest case of two-state systems with $n=2$. 
However, our results can be seen as a starting point for an extension to any value of $n$.
The density matrix is then simply a $2\times 2$  matrix that can be conveniently expanded in the basis
of the Pauli matrices $\vec{\s} \equiv (\s_1, \s_2, \s_3)  \equiv (\s_x, \s_y, \s_z)$. Using a vectorial notation, 
the expansion coefficients can be expressed in terms of the polarisation vector $\vec{P}$:
\be
\rho(t) = {1\over 2}\,(1+\vec{P}(t) \cdot \vec{\s}) = 
{1\over 2}\left(\begin{array}{cc}
1 + P_z(t) &  P_x(t) - i\,P_y(t) \\
P_x(t) + i\,P_y(t) &  1 - P_z(t)
\end{array}
\right) \,   .
\ee
Conversely, the components of the polarisation vector can be expressed in terms of density matrix entries:
\be
P_x(t)  =  2\, {\rm Re}[\rho_{12}(t)]  \,  ,  \;\;
P_y(t)  =   -2 \, {\rm Im}[\rho_{12}(t)]  \,  ,  \;\;
P_z(t)   =  \rho_{11}(t) - \rho_{22}(t) \,  .
\ee
The Hamiltonian matrix $H(t)$ can always be decomposed as
\be\label{Hdecomposition}
H(t) = {E_1(t) + E_2(t) \over 2}\, I + \D H(t) \,   ,
\ee
where $E_1(t)$ and $E_2(t)$ are the two energy eigenvalues, that, in general, are  time-dependent.  
The first term proportional to the identity does not affect the time evolution of the system. It simply yields an overall phase factor 
in the quantum state evolution. Indeed, it cancels out in the commutator of the von Neumann equation. 
We will refer to the second traceless term as the {\em effective Hamiltonian}. This can
also be expanded  in the basis of Pauli matrices:
\be\label{vecpot}
\D H(t) = {1 \over 2}\, \vec{V}(t) \cdot \vec{\s} =  
{1\over 2} \left(\begin{array}{cc}
V_z(t) & V_x(t) - i\,V_y(t) \\
V_x(t) + i\,V_y(t) &  - V_z(t)
\end{array}
\right)  \,  ,
\ee
where we  introduced the potential vector $\vec{V}(t)$ with components
\be\label{potvec}
V_x(t)  =  2\,{\rm Re}[\D H_{12}(t)]  \,  ,  \;\;
V_y(t)  =   -2\, {\rm Im}[\D H_{12}(t)]  \,  ,  \;\;
V_z(t)   =  2\D H_{11}(t) = -2 \D H_{22}(t)  \,  .
\ee
With these parameterisations of $\rho(t)$ and $H(t)$ in the basis of Pauli matrices, the von 
Neumann equation Eq.~(\ref{vonneumann2}) gets transformed into a
Larmor-like equation describing the precession of the polarisation vector around the potential vector:\footnote{This equation
was first derived in the case of nuclear magnetic resonance \cite{Bloch:1946zza} and it is usually
referred to as {\em Bloch equation} and the polarisation vector as {\em Bloch vector}.  This 
denomination is also often used in general. Notice that from $\rho^2 =\rho$, it follows that $|\vec{P}(t)|^2 =1$.
Therefore, $\vec{P}(t)$ evolves with time on a unit radius sphere, usually referred to as the {\em Bloch sphere}. 
This is not true when decoherence is  neglected, as we are assuming for the time being.}
\be\label{precession}
{\partial\vec{P}(t) \over \partial t} = \vec{V}(t) \times \vec{P}(t) \,   .
\ee
It will also prove convenient to write it explicitly in components:
\bea\label{dmeqvec}
\left( \begin{array}{ccc}
\dot{P}_x(t)  \\
\dot{P}_y(t)   \\
\dot{P}_z(t)
\end{array}\right)    & = &
\left( \begin{array}{ccc}
0 & -V_z(t) & V_y(t)  \\
V_z(t) & 0 & -V_x(t)   \\
- V_y(t) & V_x(t) & 0
\end{array}\right) \,
\left( \begin{array}{ccc}
P_x(t)  \\
P_y(t)   \\
P_z(t)
\end{array}\right)  \,  .
\eea
Notice that this equation is fully equivalent to the Schr\"{o}dinger equation (\ref{Scheq}), from where we started,
and, like the Schr\"{o}dinger equation, is basis invariant. 

In the energy eigenbasis, $\{\ket{E_1(t)},\ket{E_2(t)}\}$, the Hamiltonian matrix is diagonal and we can denote it by 
$D_H(t)$. The general decomposition in Eq.~(\ref{Hdecomposition}) gets specialised into 
\be
D_H(t) \equiv\left(
\begin{array}{cc}
E_1(t) & 0 \\
0 & E_2(t)
\end{array} \right) 
 ={E_1(t) + E_2(t) \over 2}\, I + {E_1(t) - E_2(t) \over 2} \, \sigma_3 \,  ,
\ee
so that one has for the effective Hamiltonian
\be
D_{\D H}(t) = - {1\over 2}\,\Delta E(t)\, \sigma_3 \,  ,
\ee
where we defined $\Delta E(t) \equiv E_2(t) - E_1(t)$.
Let us now consider the {\em interaction basis} $\{\ket{\lambda_1},  \ket{\lambda_2}\}$.  This basis should be considered time-independent. 
The change of basis $\{\ket{\lambda_1},  \ket{\lambda_2}\} \ra \{\ket{E_1(t)}, \ket{E_2(t)}\}$ is described by a unitary transformation 
$\hat U(t)$ such that $\ket{E_i(t)} = \hat U(t) \ket{\lambda_i}$ ($i=1,2$). The columns of the unitary matrix $U(t)$ with matrix elements
$U_{ij}(t) = \bra{\lambda_i} \hat U(t) \ket{\lambda_j}$ give the components of $\ket{E_1(t)}$ and $\ket{E_2(t)}$ in the interaction basis, since 
$\brac{\lambda_i}\ket{E_j(t)} = U_{ij}(t)$. It can be parameterised as
\be
U(t) = \left( \begin{array}{ccc}
\cos\theta(t)\,e^{i\alpha(t)} &  \sin\theta(t) \, e^{i[\alpha(t)+\varphi(t)]} \\
-\sin\theta(t) & \cos\theta(t) \, e^{i\varphi(t)}
\end{array}\right) \,  ,
\ee
where, without loss of generality, we choose the second entry in the first column to be real.\footnote{In the case that the 
Hamiltonian is time-independent, the phase $\alpha(t)$ is constant and the $\ket{E_i}$'s  are stationary states.}
On the other hand, the components of the interaction eigenstate $\ket{\lambda_j}$  in the energy eigenbasis are given by $U^\star_{j i}$
so that $\ket{\lambda_j} = \sum_i \, U^\star_{j i}(t) \ket{E_i(t)}$. We will refer to the matrix $U(t)$ as the {\em mixing matrix}
and the angle $\theta(t)$  as the {\em mixing angle}. 
In the interaction basis the effective Hamiltonian matrix is given by\footnote{Comparing this equation with the
definition of $\D H(t)$ in Eq.~(\ref{Hdecomposition}), one easily finds 
\be\label{mixang}
\alpha(t) = {\rm Arg}(H_{12}(t)) \, , \; \sin 2\theta(t) = 2 {|H_{12}(t)|\over \Delta E(t)} \,  ,  \; 
\cos 2\theta(t) =  {H_{22}(t) - H_{11}(t) \over \Delta E(t)}\,  .
\ee
 Notice that the invariance of the trace implies $E_1(t) +E_2(t) = H_{11}(t) + H_{22}(t)$ 
 and the invariance of the determinant implies
$E_1(t) E_2(t) = H_{11}(t) H_{22}(t) - |H_{12}(t)|^2$. Combining them together, one has
$\Delta E(t)^2 = (H_{22}(t) - H_{11}(t))^2 +4 |H_{12}(t)|^2$, showing
that the expressions in Eq.~(\ref{mixang}) satisfy correctly $\sin^2 2\theta(t) + \cos^2 2\theta(t) = 1$.} 
\bea\label{effectiveH}
\D H(t) & = &  U(t) \, D_{\D H}(t) \, U^\dagger(t) \\ \nonumber
&   = & -{1 \over 2}\, \D E(t) \, U(t) \, \sigma_3 \,  U^\dagger(t)  \\ \nonumber
& = &  - {1 \over 2}\, \D E(t) \, \left(
\begin{array}{cc}
\cos 2\theta(t) & -\sin 2\theta(t)\,e^{i\alpha(t)} \\
-\sin 2\theta(t)\,e^{-i\alpha(t)} & -\cos 2\theta(t) 
\end{array}
\right) \, .
\eea
From Eq.~(\ref{potvec}) one then finds:
\be
\vec{V}(t) = \D E(t) \,(\sin 2\theta(t)\cos\alpha(t), -\sin 2\theta(t)\sin\alpha(t), -\cos 2\theta(t))   \,    .
\ee
With this expression for $\vec{V}(t)$,  Eq.~(\ref{dmeqvec}) becomes
\be\label{dmeqvec4}
\left( \begin{array}{ccc}
\dot{P}_x(t)  \\
\dot{P}_y(t)   \\
\dot{P}_z(t)
\end{array}\right)     = 
\D E(t)
\left( \begin{array}{ccc}
0 & \cos 2\theta(t) & -\sin 2\theta(t)\sin\alpha(t)   \\
-\cos 2\theta(t) & 0 & - \sin 2\theta(t)\cos\alpha(t)  \\
\sin 2\theta(t) \sin\alpha(t) & \sin 2\theta(t) \cos\alpha(t)  & 0
\end{array}\right) 
\left( \begin{array}{ccc}
P_x(t)  \\
P_y(t)   \\
P_z(t)
\end{array}\right) \,  .
\ee
Notice that the mixing angle $\theta(t)$ and the phase $\alpha(t)$ determine the orientation of the potential vector $V(t)$. 
More specifically $\theta(t)$ corresponds to the polar angle and $\alpha(t)$ to the azimuthal angle in the 
three-dimensional space of traceless two-by-two matrices.

We will consider a time-dependence only in the polar angle but not in the azimuthal angle. This corresponds to take a real effective Hamiltonian
(as one can see from Eq.~(\ref{mixang})). 
If the azimuthal angle  is a constant, then it does not influence the final value of $P_{1\ra 2}$ and in that case it can be simply
set to zero.  We will be back on this assumption and comment on the possible impact of an azimuthal angle time-dependence on our results. 
Therefore, with these assumptions, the potential vector specialises into 
\be
\vec{V}(t) = \D E(t) \,(\sin 2\theta(t), 0 , -\cos 2\theta(t))   \,    ,
\ee
and the set of differential equations (\ref{dmeqvec4}) into 
\be\label{dmeqvec2}
\left( \begin{array}{ccc}
\dot{P}_x(t)  \\
\dot{P}_y(t)   \\
\dot{P}_z(t)
\end{array}\right)     = 
\D E(t)
\left( \begin{array}{ccc}
0 & \cos 2\theta(t) & 0   \\
-\cos 2\theta(t) & 0 & - \sin 2\theta(t) \\
0 & \sin 2\theta(t) & 0
\end{array}\right) 
\left( \begin{array}{ccc}
P_x(t)  \\
P_y(t)   \\
P_z(t)
\end{array}\right) \,  .
\ee

\section{Mixing angle in the general case}

It is useful to decompose the Hamiltonian operator in a time-independent part $H_0$ and
a time-dependent part $V(t)$, so that $H(t) = H_0 + V(t)$. Notice that, in general, the time-dependent part is not
necessarily a perturbation but it can be dominant or even the only term,  if $H_0 = 0$.
With this decomposition, the effective Hamiltonian in Eq.~(\ref{effectiveH}), written in the interaction basis where the states are produced and measured,
can be written as\footnote{We denote with a subscript ``$_0$'' time-independent quantities. In particular $\Delta E_0 \equiv E_{20} - E_{10}$,
where $E_{10}$ and $E_{20}$ are the eigenvalues of $H_0$.}
\be\label{effectiveHb}
\D H(t)  =  - {1 \over 2}\, \left(
\begin{array}{cc}
\D E_0 \, \cos 2\theta_0 - \D V(t)& -\D E_0 \, \sin 2\theta_0 - 2 V_{12}(t) \\
-\D E_0 \, \sin 2\theta_0 - 2 V^\star_{12}(t)& -\D E_0 \, \cos 2\theta_0 + \D V(t)
\end{array}
\right) \,  ,
\ee 
where $\D V(t) \equiv V_{11}(t) - V_{22}(t)$ comes from the time-dependent diagonal interaction terms, while
 $V_{12}(t)$ is a possible off-diagonal term originating from the presence of additional interactions of the system and that for the time being we 
 still take as complex to be general.
 For example, in neutrino physics this term could originate from non-standard interactions that, in general, are non-diagonal in the standard interaction basis. 
There are different situations one can envisage. Something that is worth to notice is that, if there is no constant  mixing angle, either 
simply because $\theta_0 =0$ or $\Delta E_0 = 0$, then a mismatch between the interaction basis and the energy eigenbasis is still 
possible if $V_{12}(t) \neq 0$. In this case the mixing angle is of course time-dependent. 
If $\D E_0 \neq 0$, we can factorise $\D E_0$ and rewrite $\D H(t)$ in the dimensionless form 
\be\label{effectiveH2}
\D H(t)  =  - {1 \over 2}\, \D E_0 \left(
\begin{array}{cc}
\cos 2\theta_0 - \D v(t)& -\sin 2\theta_0 +v_{12}(t) \\
-\sin 2\theta_0 + v_{12}^\star(t)& - \cos 2\theta_0 + \D v(t)
\end{array}
\right) \,  ,
\ee 
where we introduced the dimensionless quantities $\D v(t) \equiv \D V(t)/\D E_0$
and $v_{12}(t) \equiv -2V_{12}(t)/\D E_0$. The potential vector can then be written as
\be\label{potvecgen}
\vec{V}(t) \equiv \D E_0 \,(\sin 2\theta_0 -{\rm Re}[v_{12}(t)],  {\rm Im}[v_{12}(t)], \D v(t) -\cos 2\theta_0)   \,  .
\ee
We can now diagonalise the  effective Hamiltonian, finding time-dependent energy levels and mixing angle.
In this way we can write the density matrix equation in the vectorial form Eq.~(\ref{dmeqvec2}). The eigenvalues of the dimensionless
matrix $-2\D H(t)/\D E_0$ in Eq.~(\ref{effectiveH2}) are easily found and given by
\be
\lambda_{\pm}(t) = \pm \sqrt{[\cos 2\theta_0 - \D v(t)]^2 + |\sin 2\theta_0-v_{12}(t)|^2} \,  ,
\ee
from which one finds 
\be
\Delta E(t) = {1\over 2} \D E_0 [\lambda_+(t) -\lambda_-(t)] = \D E_0 \, \sqrt{[\cos 2\theta_0 - \D v(t)]^2 + |\sin 2\theta_0-v_{12}(t)|^2} \,  .
\ee
Let us now, as explained above, focus on the case where the azimuthal angle $\alpha(t)$ is set to zero.
We can derive the time-dependent mixing angle, considering that the diagonalising matrix $U(t)$ bringing from the interaction basis 
to the energy eigenbasis  has to satisfy Eq.~(\ref{effectiveH}). Considering a real Hamiltonian, as we discussed, this can be now rewritten 
explicitly as
\bea \nonumber
\left(
\begin{array}{cc}
\cos 2\theta_{0}- \D v(t) & -\sin 2\theta_{0} + v_{12}(t) \\
-\sin 2\theta_{0} + v_{12}(t) & -\cos 2\theta_{0} + \D v(t) 
\end{array}
\right)  & = & \left( \begin{array}{cc}
\cos \theta(t) &  \sin\theta(t)  e^{i\varphi(t)}\\
-\sin \theta(t) & \cos \theta(t) e^{i\varphi(t)}
\end{array}\right) 
\, 
\left( \begin{array}{cc}
\lambda_+(t) &  0  \\
0 & \lambda_-(t)
\end{array}\right)  \\
\,  &  &
 \left( \begin{array}{cc}
\cos \theta(t) &  -\sin\theta(t)   \\ 
 \sin \theta(t) e^{-i\varphi(t)} & \cos \theta(t) e^{-i\varphi(t)}
\end{array}\right) \, .
\eea
From first column entries, considering that $\lambda_-(t) = -\lambda_+(t)$, one has
\bea
\cos 2\theta_0 - \Delta v(t) & = & \lambda_+ (t)\, \cos 2\theta(t) \,  , \\
\sin 2\theta_0 -v_{12}(t) &  = & \lambda_+(t) \,  \sin 2\theta(t) \,  ,
 \eea
from which it immediately follows:\footnote{
This of course also implies: 
\be\label{thetat}
\cos 2\theta(t)  =  {\cos 2\theta_0 - \Delta v(t) \over \sqrt{[\cos 2\theta_0 - \D v(t)]^2 + [\sin2\theta_0 - v_{12}(t)]^2}} \,  \,   , \;\;\;\;
\sin 2\theta(t)    =  {\sin 2\theta_0 - v_{12}(t) \over  \sqrt{[\cos 2\theta_0 - \D v(t)]^2 + [\sin 2\theta_{0}-v_{12}(t)]^2}}\,  .
 \ee}
\be\label{tan2theta}
\tan 2\theta(t) = {\sin 2\theta_0 - v_{12}(t) \over \cos 2\theta_0 - \D v(t)} \,  .
\ee
With these results, we can now rewrite the potential vector as
\be
\vec{V}(t) \equiv \D E_0\,(\sin 2\theta_0 - v_{12}(t),  0 , \Delta v(t) -\cos 2\theta_0)   \,  .
\ee
Notice once more that, in the case of a real Hamiltonian, the $y$ component of the potential vector vanishes. 
If $\cos 2\theta_0$ and $\Delta v(t)$ can have equal sign, then there can be a  special time 
$t_{\rm res}$, where $\Delta v(t_{\rm res}) = \cos 2\theta_0$ and the mixing angle $\theta(t_{\rm res}) = \pm \pi/4$ \cite{Bloch:1946zza}.
Notice that the condition $\Delta v(t_{\rm res}) = \cos2\theta_0$
is equivalent to have $\Delta V(t) = \Delta E_0 \cos 2\theta_0$, i.e., it occurs when the  time-dependent 
energy splitting  equals the time-independent energy splitting, i.e., 
the time-dependent frequency is equal (times $\cos 2\theta_0$) to the natural oscillation frequency of 
the system (in the absence of time-dependent part).  Therefore, at $t_{\rm res}$ there is
a resonant behaviour, since  the two frequencies coincide and this corresponds to a maximal value $\sin^2 2\theta(t_{\rm res}) = 1$
(in particle mixing this corresponds to maximal mixing).   Moreover, at $t_{\rm res}$ the energy level spacing, $\Delta E(t)$, reaches a minimum. 
We will  refer to $t_{\rm res}$ as the resonance time. One expects maximum transition rate around this time. 

\section{Time-independent Hamiltonian operator}

Let us consider the simplest case of a time-independent Hamiltonian operator. This corresponds to have a constant mixing angle $\theta_0$ and 
a constant potential vector
\be 
\vec{V}_0 \equiv \D E_0 \,(\sin 2\theta_0, 0, -\cos 2\theta_0)   \,  .
\ee  
The solution of Eq.~(\ref{dmeqvec2}) can be derived straightforwardly.
Suppose that at the initial time, $t=0$, the state is produced in the interaction eigenstate $\ket{\lambda_1}$.
This implies that at the initial time the density matrix is simply given by
\be\label{initialdm}
\rho(0) = \left( \begin{array}{ccc}
1 &  0 \\
0 & 0
\end{array}\right) \,  ,
\ee
corresponding to $\vec{P}(0) = (0,0,1)$. Therefore, at the initial time, the 
polarisation vector lies along the $z$-axis in the positive direction. 
The solution of Eq.~(\ref{dmeqvec2}) can be found easily in a standard way, finding eigenvalues and eigenvectors of the coupling matrix
and writing the solution as a linear combination of the eigenvectors with time-dependent exponential factors containing the eigenvalues in the 
exponents.  Let us follow an alternative geometrical procedure considering the vectorial form of the equation. The solution is a time-evolving polarisation
vector precessing anti-clockwise around $\vec{V}_0$ with angular velocity $\omega =|\vec{V}_0| = \D E_0$.  This can be seen explicitly in rather a simple way. 
We make a change of variables, $\vec{P}(t) \ra \vec{P}'(t)$, performing a rotation around the $y$ axis by an angle $2\theta_0$, 
in a way that the $z'$-axis is anti-parallel to $\vec{V}_0$.  
Explicitly, in this new coordinate system, one has $\vec{P}(t) \ra \vec{P}'(t)= R_y(2\theta_0)\,\vec{P}(t) $, where the rotation matrix 
\be
R_y(2\theta_0) =
\left( \begin{array}{ccc}
\cos 2\theta_0 &  0  & +\sin 2\theta_0 \\ 
0 & 1 & 0 \\
-\sin 2\theta_0 & 0 &  \cos 2\theta_0
\end{array}\right) \,  .
\ee
One can verify that indeed $\vec{V}_0' = R_y(2\theta_0) \,\vec{V}_0 = \D E_0 \, (0, 0, -1)$.\footnote{Notice that in terms of
quantum states, in this new coordinate system the $z'$ axis identifies a polarisation vector describing the energy eigenstates: the positive 
direction corresponds to $\ket{E_1}$ and the negative direction to $\ket{E_2}$. This is clear since 
if $\vec{P}'$ is parallel to $\vec{V}_0$, then from the Block equation one has that $\dot{P}' =0$, signalling that the system
is in a stationary state.}
We can then write $\vec{P}(t) = R^T_y(2\theta_0) \, \vec{P}'(t)$ and the system of coupled differential equations (\ref{dmeqvec2}) 
now becomes,  in the new primed variables, a system of decoupled differential equations:
\bea\nonumber
\left( \begin{array}{ccc}
\dot{P}'_x(t)  \\
\dot{P}'_y(t)   \\
\dot{P}'_z(t)
\end{array}\right)    & = &
\D E_0 \, 
R_y\, 
\left( \begin{array}{ccc}
0 & \cos 2\theta_0 & 0  \\
-\cos 2\theta_0 & 0 & - \sin 2\theta_0   \\
0 & \sin 2\theta_0 & 0
\end{array}\right) \, R_y^T \,
\left( \begin{array}{ccc}
P'_x(t)  \\
P'_y(t)   \\
P'_z(t)
\end{array}\right) \,  , \\ \nonumber
& = &
\D E_0\,  
\left( \begin{array}{ccc}
0 & 1 & 0  \\
-1 & 0 & 0   \\
0 & 0 & 0
\end{array}\right) \, 
\left( \begin{array}{ccc}
P'_x(t)  \\
P'_y(t)   \\
P'_z(t)
\end{array}\right)  \,  .
\eea
Considering moreover that $\vec{P}'(0) = R_y(2\theta_0)\,\vec{P}(0) = (\sin 2\theta_0, 0 ,\cos 2\theta_0)$, 
one finds the following easy solution:
\bea\label{timeindsol}
P'_x(t) & = & \sin 2\theta_0\,\cos\left(\D E_0\,t \right) \,  ,\\ \nonumber
P'_y(t) & = & -\sin 2\theta_0\,\sin\left(\D E_0\,t \right) \,  ,\\ \nonumber
P'_z(t) & = & \cos 2\theta_0 \,  .
\eea
Notice that $\vec{P}'(t)$ precedes around $\vec{V}_0$ in a way that both the parallel component, $P'_z=\cos 2\theta_{0}$, and the 
length of the transversal component, $P'_{\bot} =\sqrt{P_x^{'2}+P_y^{'2}}= \sin2\theta_{0}$, remain constant.
Performing the inverse transformation, $\vec{P}(t) = R_y^T \, \vec{P}'(t)$, explicitly 
\be
\left( \begin{array}{ccc}
P_x(t)  \\
P_y(t)   \\
P_z(t)
\end{array}\right) = \left( \begin{array}{ccc}
\cos 2\theta_0 &  0  & -\sin 2\theta_0  \\ 
0 & 1 & 0 \\
\sin 2\theta_0 & 0 &  \cos 2\theta_0 
\end{array}\right) \, \left( \begin{array}{ccc}
\sin 2\theta\,\cos\left(\Delta E_0\,t\right) \\
-\sin 2\theta\,\sin\left(\Delta E_0\,t \right)   \\
\cos 2\theta_0
\end{array}\right) \,  ,
\ee
one finally arrives to the solution for the $z$ component 
\bea\label{Pzt}
P_z(t) 
           & = & 1 - 2\sin^2 2\theta_0\, \sin^2\left({\Delta E_0\,t \over 2}\right)  \,  .
\eea
Since $\rho_{11} = (1+ P_z)/2$ and $\rho_{22} = (1-P_z)/2$, one also has
\be\label{rhoentries}
\rho_{11}(t)  =  1 - \sin^2 (2\theta_0)\, \sin^2\left({\Delta E_0\,t \over 2}\right) \,  ,  \;\;
\rho_{22}(t)  =   \sin^2 (2\theta_0)\, \sin^2\left({\Delta E_0\,t \over 2}\right) \,  ,
\ee
giving the probabilities at time $t$ to find the system in the state $\ket{\lambda_1}$ and $\ket{\lambda_2}$, respectively.
In Fig.~1 we show a graphical representation of the simple motion of the polarisation vector in the time-independent case for some generic
$\theta_0$.
\begin{figure}[t]
\begin{center}
\psfig{file=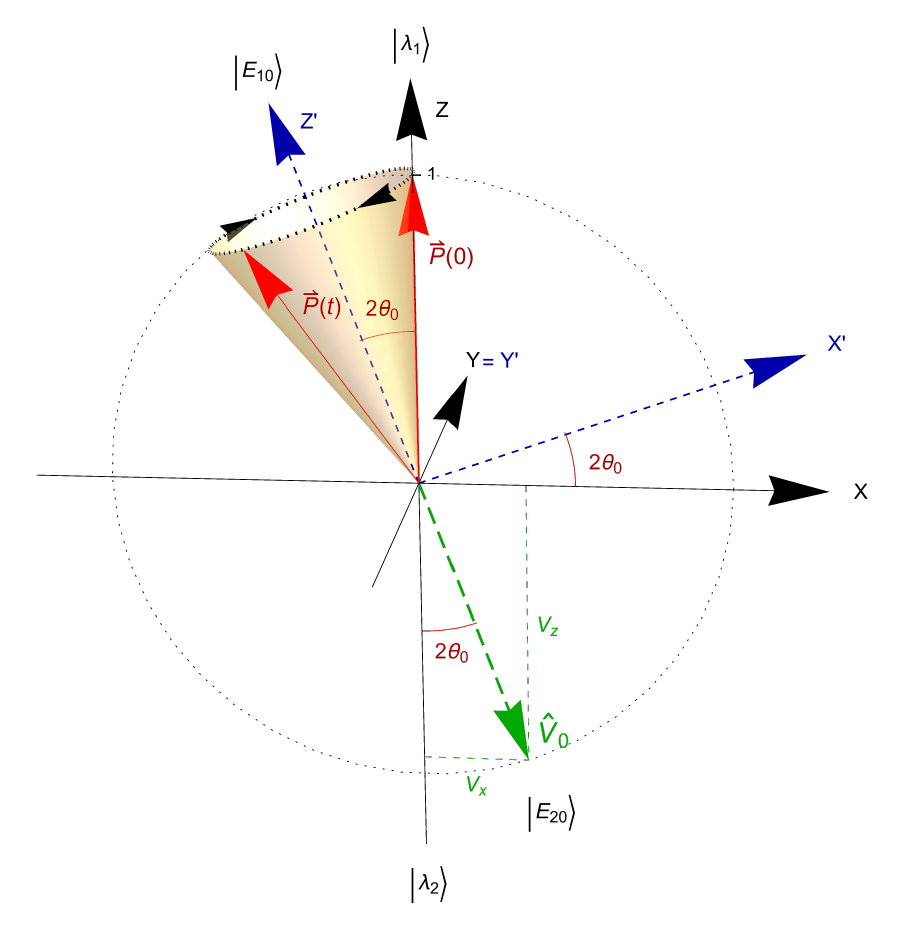,scale=0.7} 
\end{center}\vspace{-5mm}
\caption{Graphical representation of the motion of the polarization vector in the case of time-independent Hamiltonian, corresponding
to a constant potential vector $\hat{V}_0 \equiv \vec{V}_0 /\D E_0 = (\sin 2\theta_0, 0, -\cos 2\theta_0)$ for some generic $\theta_0$.
The dashed line circle is the projection of the Bloch sphere on the $x-z$ plane.}
\end{figure}

\subsection{Solar neutrino oscillations in vacuum}

An important application is provided by {\em neutrino oscillations in vacuum} \cite{Giunti:2007ry}. 
In this case the interaction eigenstates are the weak eigenstates
$\ket{\nu_\alpha}$ ($\alpha=e,\mu,\tau$) and the energy eingestates coincide with the mass eigenstates $\ket{\nu_i}$ ($i=1,2,3$).
In general, one should consider medium effects. However, in many cases these can be neglected and one can simply consider
neutrino oscillations in vacuum. For example, neutrino oscillations in vacuum well describe solar neutrino oscillations at energies 
$E \ll 2\,{\rm MeV}$, since in this case matter effects can be neglected. Moreover, approximately, solar neutrino oscillations
can be described by two-neutrino oscillations. In this case we have that the initial
state $\ket{\lambda_1} =\ket{\nu_e}$,  while $\ket{\lambda_2} =\ket{\nu_{\mu+\tau}}$, where $\nu_{\mu+\tau}$ is
a linear combination of muon and tauon neutrino states set by the atmospheric neutrino mixing angle.   
For neutrinos of a given momentum $E \equiv |\vec{p}|$, in the ultra-relativistic  limit, one has $E_{i0} \simeq E + m^2_i/(2E)$, so that
simply  $\D E_0 \simeq  \Delta m^2_{12}/2E$, with $\Delta m^2_{12} \equiv m^2_2 -m^2_1$. 
The mixing angle $\theta_0$ is in this case corresponding to the solar mixing angle $\theta_{12}$.
Neutrino oscillation experiments find $\Delta m^2_{12} \simeq  7.5 \times 10^{-5}\,{\rm eV}^2$ and $\theta_{12} \simeq 33.8^\circ$
\cite{Esteban:2024eli}. The vector potential is then given by
\be
\vec{V}_0 \simeq {\Delta m^2_{12} \over 2E} \,(\sin 2\theta_{12}, 0, -\cos 2\theta_{12})   \,    .
\ee
The density matrix element $\rho_{22}$ gives the usual appearance probability \cite{Pontecorvo:1967fh,Gribov:1968kq}:
\be\label{vacuumapp}
P_{\nu_e \ra \nu_{\mu+\tau}} = \rho_{22} = \sin^2 2\theta_{12}\, \sin^2\left({\Delta m^2_{12}\,{\ell_\odot} \over 4E}\right) \,  ,
\ee
where $\ell_\odot \simeq c t$ is the Earth-Sun distance given, on average, by the astronomical unit. 

\section{Adiabatic transitions}

Let us now consider the case when the Hamiltonian operator explicitly depends on time.  This implies that the potential vector, defined in Eq.~(\ref{vecpot}), 
is now also depending on time.  Even though the density matrix equation in vectorial form, Eq.~(\ref{dmeqvec2}), remains unchanged, it is now a 
set of linear differential equations with time-dependent coefficients. This makes a derivation of its solution non-trivial at all.  

We can again follow the same procedure as in the time-independent case to solve Eq.~(\ref{dmeqvec2}).
We can again introduce a (this time time-dependent) rotation matrix
\be
R(t) \equiv R_y(2\theta(t)) =
\left( \begin{array}{ccc}
\cos 2\theta(t) &  0  & \sin 2\theta(t) \\ 
0 & 1 & 0 \\
-\sin 2\theta(t) & 0 &  \cos 2\theta(t)
\end{array}\right) \,  ,
\ee
such that 
$\vec{V'}(t)= R(t) \, \vec{V}(t) = \Delta E(t)(0,0,-1)$ is now antiparallel to the new $z'(t)$ axis.
Making the change of variables $\vec{P}(t) \ra \vec{P}'(t)= R(t)\,\vec{P}(t)$, Eq.~(\ref{dmeqvec2}) becomes:
\be\label{rotated}
\left( \begin{array}{ccc}
\dot{P}'_x(t)  \\
\dot{P}'_y(t)   \\
\dot{P}'_z(t)
\end{array}\right)     = \left[
\D E(t)\, 
\left( \begin{array}{ccc}
0 & 1& 0  \\
- 1 & 0 & 0  \\
0 & 0 & 0
\end{array}\right) - R(t)\dot{R}^T(t) \right]\, 
\left( \begin{array}{ccc}
P'_x(t)  \\
P'_y(t)   \\
P'_z(t)
\end{array}\right) \,  .
\ee
Introducing the {\em adiabaticity parameter},
\be\label{adpar}
\gamma(t) \equiv {1\over 2}\left|{\Delta E(t) \over \dot{\theta}(t)} \right| \,  ,
\ee
Eq.~(\ref{rotated}) can be compactly rewritten as\footnote{One can indeed easily check that
\be
R(t)\dot{R}^T(t)  =  -2\dot{\theta}(t)\, 
\left( \begin{array}{ccc}
0 & 0 & 1  \\
0 & 0 & 0  \\
1 & 0 & 0
\end{array}\right) \,  .
\ee
} 
\be
\left( \begin{array}{ccc}
\dot{P}'_x(t)  \\
\dot{P}'_y(t)   \\
\dot{P}'_z(t)
\end{array}\right)     =
\D E(t)\, 
\left( \begin{array}{ccc}
0 & 1& {1 \over \gamma(t)}\,{\rm sgn}(\Delta E/\dot{\theta})\\
- 1 & 0 & 0  \\
{1 \over \gamma(t)}\, {\rm sgn}(\Delta E/\dot{\theta}) & 0 & 0
\end{array}\right)
\left( \begin{array}{ccc}
P'_x(t)  \\
P'_y(t)   \\
P'_z(t)
\end{array}\right) \,  .
\ee
If we take the {\em adiabatic limit}, holding when $\gamma(t) \gg 1$ at all times in a way that the term containing the time derivative can be neglected, 
the set of differential equations simplifies into:
\bea
\dot{P}'_x(t)  & = &  \Delta E(t) \,  P'_y(t)  \, , \\ \;\;
\dot{P}'_y(t)  &  = &  -\Delta E(t) \,  P'_x(t)  \,  , \\  \;\;
\dot{P}'_z(t)  & = &  0   \,  .
\eea
Geometrically, the adiabatic limit is equivalent to assume that the polarisation vector precedes around the potential vector at all times.  
Let us again assume that at the initial time, $t=0$, the state is produced in the interaction eigenstate $\ket{\lambda_1}$
and that the density matrix is given by Eq.~(\ref{initialdm}) corresponding to $\vec{P}(0) = (0,0,1)$ and
$\vec{P}'(0)= R(0)\,\vec{P}(0) = (\sin 2\theta(0),0,\cos 2\theta(0))$. 
In this way, one straightforwardly finds the solution in the time-dependent rotated basis generalising Eq.~(\ref{timeindsol}):
\bea
P'_x(t) & = & \sin 2\theta(0)\,\cos\left(\int_0^t\,dt'\,\Delta E(t') \right) \,  , \\ \nonumber 
P'_y(t) & = & -\sin 2\theta(0)\,\sin\left(\int_0^t\,dt'\,\Delta E(t') \right) \,   , \\ \nonumber
P'_z(t) & = & \cos 2\theta(0) \,  .
\eea
Using then $\vec{P}(t) = R^{\rm T}(t) \, \vec{P}'(t)$, i.e.,
\be
\left( \begin{array}{ccc}
P_x(t)  \\
P_y(t)   \\
P_z(t)
\end{array}\right) = \left( \begin{array}{ccc}
\cos 2\theta(t) &  0  & -\sin 2\theta(t)  \\ 
0 & 1 & 0 \\
+\sin 2\theta(t)& 0 &  \cos 2\theta(t) 
\end{array}\right) \, \left( \begin{array}{ccc}
\sin 2\theta(0)\,\cos\left(\int_0^t\,dt'\,\Delta E(t') \right)  \\
-\sin 2\theta(0)\,\sin\left(\int_0^t\,dt'\,\Delta E(t') \right)   \\
\cos 2\theta(0)
\end{array}\right) \, ,
\ee
one arrives to the solution in the interaction basis:
\be
P_z(t) = \cos 2\theta(t) \, \cos 2\theta(0) + 
\sin 2\theta(t)\,\sin 2\theta(0)\,\cos\left(\int_0^t\,dt'\,\Delta E(t')\,t' \right) \,  .
\ee
We can again immediately derive $\rho_{11}(t) = [1+ P_z(t)]/2$ and $\rho_{22}(t) = [1-P_z(t)]/2$:
\be\label{probs}
P_{1\ra 1}(t)  =   \rho_{11}(t)  \hspace{-1mm} =  {1 \over 2} + {1\over 2}\cos 2\theta(t)\cos 2\theta(0)  +  {1\over 2}\,\sin 2\theta(t)\sin 2\theta(0)\cos\left( \int_0^t \,dt' \Delta E(t') \right) ,   
\ee
\be\label{conversionprob}
P_{1\ra 2}(t)   =  \rho_{22}(t)  =   {1 \over 2} - {1\over 2}\cos 2\theta(t)\cos 2\theta(0)  -  
{1\over 2}\,\sin 2\theta(t)\sin 2\theta(0)\cos\left( \int_0^t \,dt' \Delta E(t') \right)  \,  .
\ee
These expressions give the probabilities to find the system in the state $\ket{\lambda_1}$ and $\ket{\lambda_2}$ at the time $t$, respectively.

\subsection{MSW effect}

An important neutrino physics application  is provided by the {\em MSW effect} \cite{Wolfenstein:1977ue,Mikheyev:1985zog}, 
well describing electron solar neutrino survival and disappearance
probabilities also at energies $E \gtrsim 2\,{\rm MeV}$.  As in the vacuum case, we have again $\D E_0 \simeq  \Delta m^2_{12}/2E$, 
with $\Delta m^2_{12} \equiv m^2_2 -m^2_1$ and $\theta_0 = \theta_{12}$.
Solar neutrinos are produced in the centre of the Sun and propagate outside in a straight line. Despite they do not undergo collisions changing their
momentum, one still has to take into account forward scatterings, a self-energy contribution originating from the same electroweak 
interactions that produced them. In this way a term $V(t)$ is generated. This is diagonal
in the interaction basis, so that  $V_{12}(t) =0$. The diagonal term, usually referred to as effective potential in neutrino physics, is given by 
$\Delta V(t) = {\sqrt{2}\,G_F\,n_e(r(t))}$, where $n_e(r(t))$ is the electron number density inside the sun, approximately given by 
\be
n_e(r) = n_e(0)\, e^{-8.4 {r\over R_\odot}} \,  ,
\ee
with the central value $n_e(0) \simeq 5 \times 10^{31}\,{\rm m}^{-3}$ \cite{Laber-Smith:2024hbc}.
As in the vacuum case, we can approximate $r(t) \simeq c t$. The Eq.~(\ref{effectiveH2}) gets now specialised into
\be\label{effectiveH2b}
\D H(t)  =  - {\D m^2_{12} \over 4 E}\, \left(
\begin{array}{cc}
\cos 2\theta_{12} - v_e(t)& -\sin 2\theta_{12}  \\
-\sin 2\theta_{12} & - \cos 2\theta_{12} + v_e(t)
\end{array}
\right) \,  ,
\ee 
where we used
\be\label{Dvt}
\D v(t) = v_e(t) \equiv {2 E \over \Delta m^2_{12}}\,\sqrt{2}G_{\rm F}\,n_e(r) \,  .
\ee
Since we detect neutrinos at the Earth's surface, at a huge distance from
the Sun, the cosine in  Eqs.~(\ref{probs}) and (\ref{conversionprob}) averages to zero.
Moreover, we can replace $\theta(t)$ with the vacuum mixing angle $\theta_{12}$
and in this way one obtains for the survival and appearance probabilities 
\bea\label{probs2}
P_{\nu_e \ra \nu_e} \hspace{0mm}  & = &  {1 \over 2} + {1\over 2}\cos 2\theta_{12}\cos 2\theta(0)  \,  ,  \\ \label{conversionprob2}
P_{\nu_e \ra \nu_{\mu+\tau}} \hspace{0mm} & = &    {1 \over 2} - {1\over 2}\cos 2\theta_{12}\cos 2\theta(0)  \,  .
\eea
For $E \gg 2\,{\rm MeV}$, one has $\Delta v(0) \gg 1$, and from Eq.~(\ref{thetat}), one can see that
$\theta(0) \ra -\pi/2$ and one obtains the limits: 
\bea
P_{\nu_e \ra \nu_e} & = & \rho_{11} \ra \sin^2 \theta_{12} \simeq  0.31\\
P_{\nu_e \ra \nu_{\mu+\tau}} & = & \rho_{22} \ra \cos^2 \theta_{12} \simeq 0.69\,   ,
\eea 
where we used $\theta_{12} \simeq 33.8^\circ$ \cite{Esteban:2024eli}.
In the four panels of Fig.~2, we show a graphical representation of the polarisation vector 
and potential vector in the adiabatic case for the MSW effect for solar neutrinos with $E = 10\,{\rm MeV}$, 
at four particular times, as described in the caption, corresponding to four different values of the variable $\xi \equiv r/R_\odot$.
\begin{figure}[t]
\begin{center}
\psfig{file=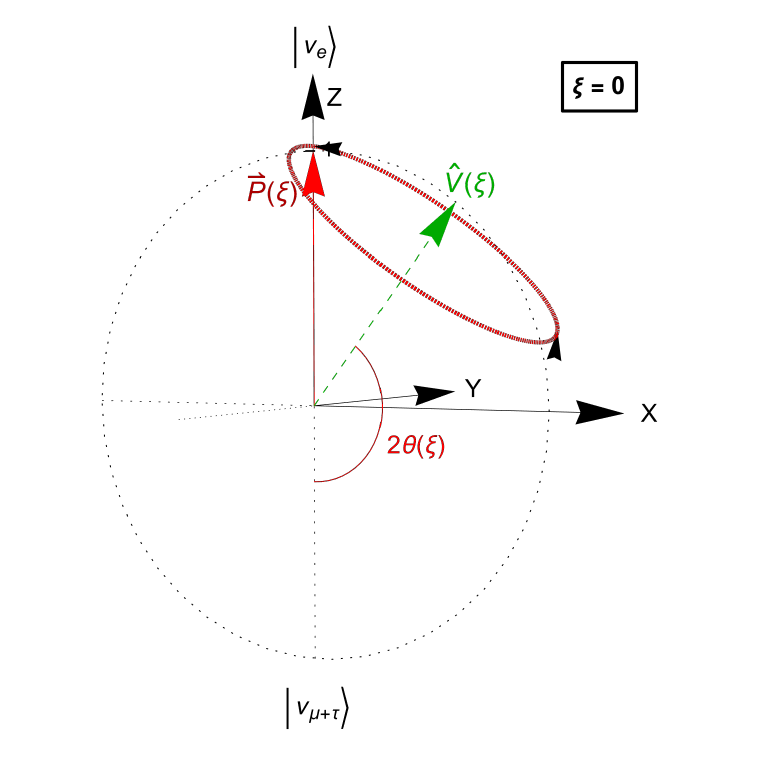,scale=0.4} \hspace{5mm} 
\psfig{file=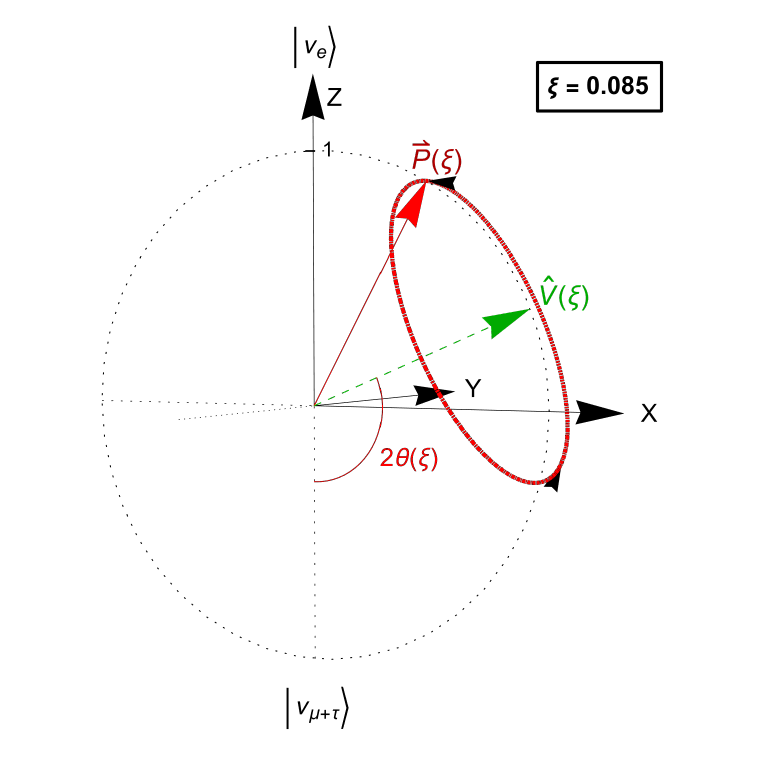,scale=0.4} \\  \vspace{0mm}
\psfig{file=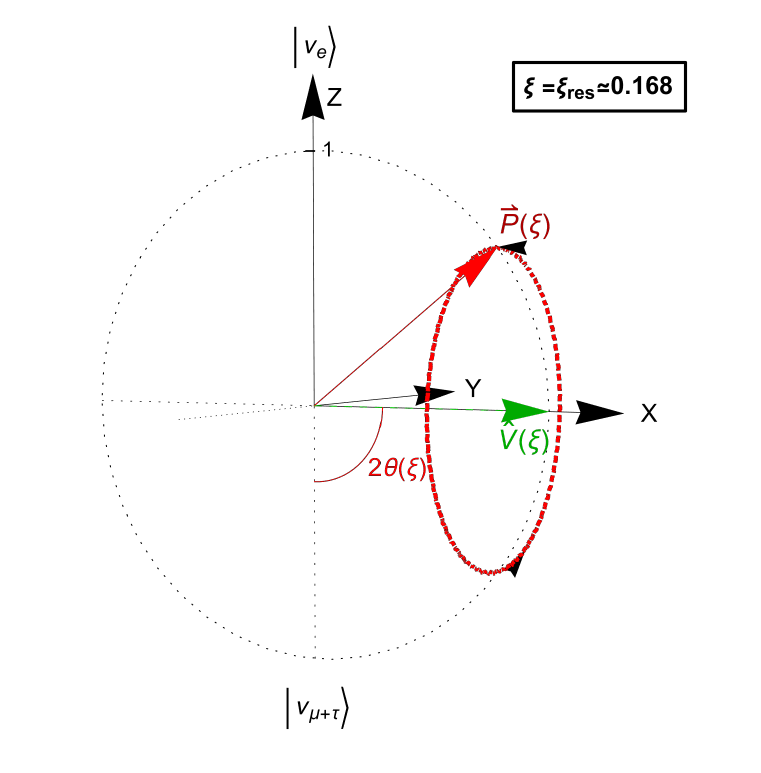,scale=0.4} \hspace{5mm} 
\psfig{file=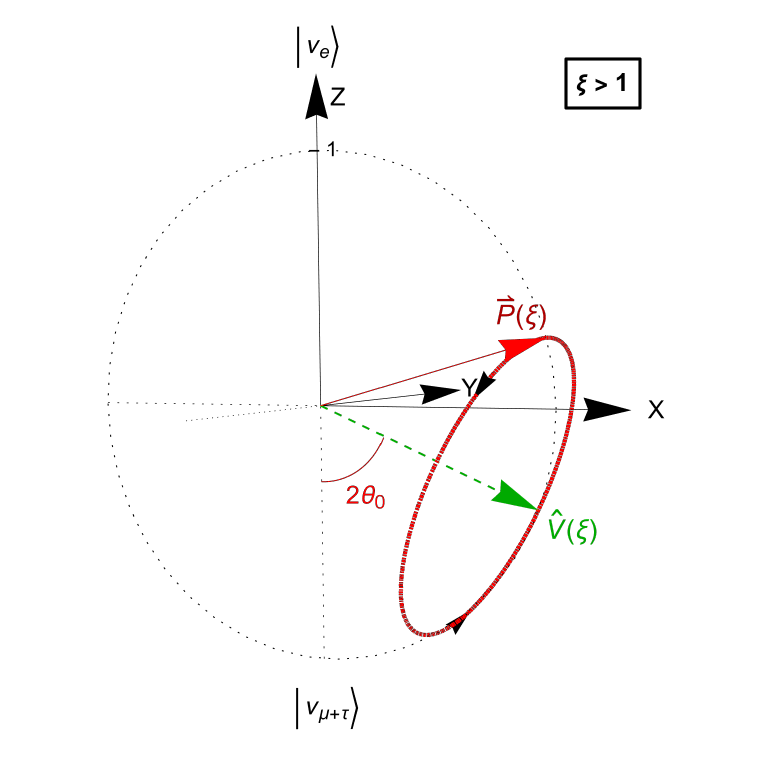,scale=0.4}  
\end{center} \vspace{-12mm}
\caption{Representation of the polarisation and normalised potential vector 
$\hat V(t) \equiv \vec{V}(t)/\Delta E(t)$ in the case of the solar MSW effect  with 
$E = 10\,{\rm MeV}$ for the four indicated values of $\xi \equiv r/R_\odot$.  {\em Top left}: $t=r=0$, in this case one has $\Delta v(0) \simeq1.576$ 
and $\theta(0) \simeq 72^\circ$.
{\em Top right}: $\xi\simeq 0.085$, implying $\Delta v(\xi)\simeq 0.77$ and $\theta(\xi)\simeq 56^\circ$;  
{\em Bottom left}: resonance, $\xi_{\rm res}\simeq 0.169$, implying $\Delta v(\xi_{\rm res}) = \cos 2\theta_{12} \simeq 0.381$
and $\theta(\xi_{\rm res}) = 90^\circ$; {\em Bottom right}: $\xi \gg \xi_{\rm res}$, vacuum oscillations are recovered. 
In each panel, the dashed circle is the projection of the Bloch sphere on the $x-z$ plane.
}
\end{figure}
In the last panel, for $\xi > 1$, vacuum oscillations are recovered.\footnote{However, notice that
for very large $\xi$  neutrino oscillations fade away, since neutrino 
mass eigenstates propagate to a slightly different speed inducing decoherence. For this reason, when 
reaching the Earth, neutrinos are detected as mass eigenstates, not as interaction eigenstates.}
Notice that from Eqs.~(\ref{conversionprob2}) one also recovers the appearance probability in the
vacuum case in Eq.~(\ref{vacuumapp}) by setting $\theta(0) = \theta_{12}$ and averaging to $1/2$
the second factor.

In general, Eqs.~(\ref{probs2}) and (\ref{conversionprob2}) can be evaluated for any generic value of $\theta(0)$ between
the vacuum limit $\theta(0) =\theta_{12}$ and the asymptotic limit $\theta(0) =\pi/2$ as a function of the energy. 
In Fig.~3 we show the result for the disappearance probability 
$P_{\nu_e \ra \nu_e}$ as a function of the neutrino energy $E$. 
One can see how this nicely interpolates between the vacuum limit at low energies and the MSW limit 
for $\theta(0) = \pi/2$ at high energies. 
\begin{figure}[t]
\begin{center}
\psfig{file=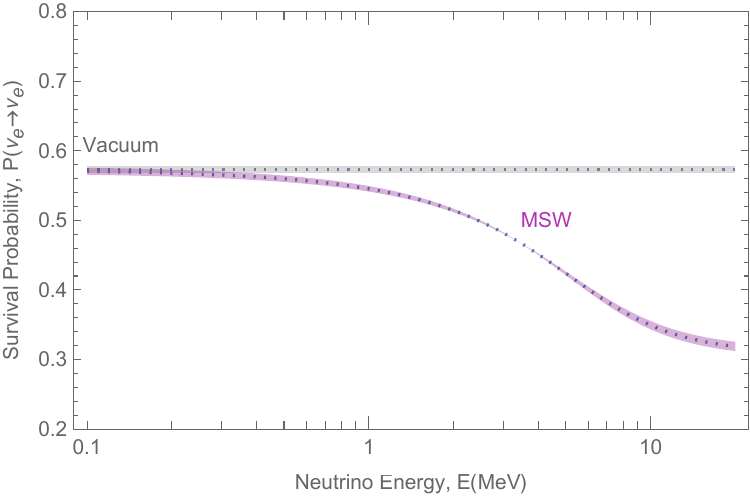,scale=0.8} 
\end{center}
\caption{Survival probability of solar neutrinos as a function of the energy $E$: vacuum (grey), MSW (purple).}
\end{figure}

\section{Non-adiabatic regime} 

Let us now discuss the non-adiabatic regime in the case of a time-dependent Hamiltonian. 
A common procedure is to start from Eq.~(\ref{rotated}) retaining the term containing the time
derivative of the matrix $R$ and, therefore, of the mixing angle. 

However, here we derive a general solution starting directly from the unrotated Eq.~(\ref{dmeqvec}) that it is convenient to recast, 
explicitly, as:
\bea\label{unrotated}
\dot{P}_x(t)  & = & -V_z(t)\, P_y(t)  \\ \nonumber
\dot{P}_y(t)  & = &   V_z(t)\,P_x(t) - V_x(t)\,P_z(t) \\ \nonumber
\dot{P}_z(t)  & = & V_x(t) \, P_y(t)  \,   .
\eea
Let us now perform the following simple change of variables $\{P_x(t), P_y(t)\} \ra \{\widetilde{P}_x(t),\widetilde{P}_y(t)\}$:
\be\label{pxtpyt}
P_x(t)  =  {\widetilde{P}_x(t) + \widetilde{P}_y(t)\over \sqrt{2}}  \, , \;\; 
P_y(t)  =  -i{\widetilde{P}_x(t) - \widetilde{P}_y(t)\over \sqrt{2}} 
\ee
After this change, the first two equations (\ref{unrotated}) become
\bea\label{unrotatedb}
\dot{\tilde{P}}_x(t)  & = & + i V_z(t)\, \widetilde{P}_x(t) -{i\over \sqrt{2}} V_x(t) \,P_z(t)  \\ \nonumber
\dot{\tilde{P}}_y(t)  & = &   - i V_z(t)\,\widetilde{P}_y(t) + {i\over \sqrt{2}}  V_x(t)\,P_z(t) \,  . 
\eea
From these equations we can write $\widetilde{P}_x(t)$ and $\widetilde{P}_y(t)$ in terms of the unknown $P_z(t)$:
\bea
\widetilde{P}_x(t) & = & \widetilde{P}_x(0)\, e^{+i \int_{0}^t dt' V_z(t')} - {i\over \sqrt{2}}\int_{0}^t dt' \, V_x(t') P_{z}(t') \, e^{i\int_{t'}^t dt'' V_z(t'')}\,  , \\
\widetilde{P}_y(t) & = & \widetilde{P}_y(0)\, e^{-i \int_{0}^t dt' V_z(t')} + {i\over \sqrt{2}}\int_{0}^t dt' \, V_x(t') P_{z}(t') \, e^{-i\int_{t'}^t dt'' V_z(t'')} \,  .
\eea
Since $\vec{P}(0)=(0,0,1)$, one also has $\widetilde{P}_x(0) = \widetilde{P}_y(0) = 0$. Switching back to the variables 
$P_x(t)$ and $P_y(t)$ given by the Eqs.~(\ref{pxtpyt}), one finds:   
\bea\label{solutionPxPy}
P_x(t) & = &  \int_{0}^t dt' \, V_x(t') P_{z}(t') \, \sin\left(\int_{t'}^t dt'' V_z(t'') \right)\,  , \\
P_y(t) & = &   -\int_{0 }^t dt' \, V_x(t') P_{z}(t') \, \cos\left(\int_{t'}^t dt'' V_z(t'') \right) \,  .
\eea
One still has to solve the following  integro-differential equation for $P_z(t)$:
\be\label{Pzderivative}
{dP_z(t) \over dt}   =  -V_x(t) \, \int_{0 }^t dt' \, V_x(t') P_{z}(t') \, \cos\left(\int_{t'}^t dt'' V_z(t'') \right)  \,  .
\ee 
In the approximation $P_z \simeq 1$, the solution is straightforward and one finds
\be\label{appearance}
P_{1\ra 2}(t) =\rho_{22}(t) = {1 - P_z(t) \over 2} = {1\over 2}\, \int_{0 }^t dt' \, V_x(t')  \int_0^{t'} dt'' \, V_x(t'') \, \cos\left(\int_{t''}^{t'} dt''' V_z(t''') \right) \,  . 
\ee
In the case of time-independent Hamiltonian one immediately recovers the solution Eq.~(\ref{rhoentries}):
\be
P_{1\ra 2}(t) = \rho_{22}(t)  =  {1\over 2}\sin^2 2\theta_0\,\int_0^y dy' \int_0^{y'} dy'' \, \cos(y'-y'')   
                     =  \sin^2 2\theta_0\,\sin^2\left({\Delta E_0\,t \over 2}\right)  \,  ,
\ee
where in the first integral $y \equiv \Delta E_0 \, t$ and we approximated $\cos (2\theta_0) \simeq 1$ to be consistent 
with $P_z \simeq 1$, an approximation valid for small mixing angles.

\subsection{LZSM approximation}

The well known LZSM approximation describing the crossing probability between the two energy eigenstates
is given by \cite{Landau:1932wdt,Zener:1932ws,Stuck1932,Majorana:1932ga} 
\be\label{LZ}
P_{\rm c} = e^{-{\pi \over 2}\, \gamma_{\rm res}} \,  ,
\ee
where $\gamma_{\rm res} = \gamma(t_{\rm res})$. One can easily find from Eq.~(\ref{adpar})
\be\label{resadpar}
\gamma_{\rm res} = \sin^2 2\theta_0 \, \left| {\Delta E_0 \over \D \dot{v}(t_{\rm res})}\right| \,  .
\ee
It is obtained assuming $v_{12}(t) =0$ and in the linear approximation 
$\Delta v \simeq \Delta v(t_{\rm res}) + \dot{\Delta}v(t_{\rm res}) (t-t_{\rm res})$.\footnote{In the Appendix
we review the original derivation by Zener \cite{Zener:1932ws} and a derivation \cite{pokrovskykhalatnikov,dykhne,Pokrovsky:2010zz}
that extends that one by Landau \cite{Landau:1932wdt}.}
For small mixing angles ($\theta_0 \ll 1$), it implies a conversion probability
\be
P_{1\ra 2}(t_{\rm f}) = 1 - e^{-{\pi \over 2}\, \gamma_{\rm res}} \,  .
\ee
Let us show that the expression in Eq.~(\ref{appearance}) with $t=\infty$ 
reproduces the LZSM formula Eq.~(\ref{LZ}) at first order in the adiabaticity parameter
within a proper set of approximations. As we have seen, at the resonance, the $z$ component of the potential vanishes: $V_z(t_{\rm res})=0$.
Away from the resonance the argument of the cosine is rapidly oscillating and the third integral would average to zero so that the main assumption is that
the dominant contribution comes from a narrow interval around the resonance. 
If we assume that the resonance is crossed quickly enough that the variation of $V_{z}(t\sim t_{\rm res})$ can be neglected, 
we can expand $V_{z}(t)$ at first order around the resonance. In this way we can replace 
$V_{z}(t''') \simeq \D E_0\,\dot{\D v}(t_{\rm res})\,(t'''-t_{\rm res})$ in the third integral, obtaining:
\be
P_{1\ra 2}^{\rm f}  = {1\over 2}\, \int_{0 }^\infty dt' \, V_x(t')  \int_0^{t'} dt'' \, V_x(t'') \, 
\cos\left(\D E_0\,\dot{\D v}(t_{\rm res})\,\left[ {(t'-t_{\rm res})^2 - (t''-t_{\rm res})^2 \over 2}\right]\right) \,  . 
\ee
We can then change both variables of integrations, $t' \ra \tau'  = t'\, \sqrt{\D E_0\,\dot{\D v}(t_{\rm res})/2}$ and 
$t'' \ra \tau''  = t''\, \sqrt{\D E_0\,\dot{\D v}(t_{\rm res})/2}$. We can also
take outside the integrals both  $V_{x}(t')$ and $V_x(t'')$, treating them as constant evaluating at the resonance (if $v_{12} =0$, 
this would be exact since in that case simply $V_x = \Delta E_0 \, \sin 2\theta_0$). In this way we obtain
\be
P_{1\ra 2}^{\rm f}  = {V_x^2(t_{\rm res})\over 2\,\D E_0\,\dot{\D v}(t_{\rm res})}\, \int_{0 }^\infty d\t' \,  \int_0^{\infty} d\t'' \, 
\cos\left(\left[ {(\t'-\t_{\rm res})^2} - {(\t''-\t_{\rm res})^2}\right]\right) \,  ,
\ee  
where we extended also the integration in $t''$ to $\infty$ and divided by a factor 2 thanks to the exchange symmetry between 
$\tau'$ and $\tau''$. 
We can then factorise the integrations on $t'$ and $t''$ by using the simple trigonometric relation $\cos (\a - \b) = \cos\a \cos\b + \sin\a\sin\b$
and also make a further approximation replacing the lower extremes of integrations from $0$ to $-\infty$. In this way one obtains
Fresnel integrals of the kind $\int_{-\infty}^{+\infty} d\tau \tau^2 = \sqrt{\pi/4}$, finally leading to the expected LZSM formula
for $\gamma_{\rm res} \ll 1$:
\be
P_{1\ra 2}^{\rm f}  \simeq {\pi\over 2}\,\gamma_{\rm res} \,  .
\ee 
This shows that the expression Eq.~(\ref{appearance}) correctly reproduces, within the appropriate limit and set of assumptions,
the correct LZSM limit.

\subsection{Extremely non-adiabatic limit}

Consider the case when the time-dependent term in the Hamiltonian, $\Delta V(t)$, is non-vanishing in some non interval of time
$\Delta t = [t_{\rm on},t_{\rm off}]$, where $t_{\rm on}$ can also coincide with the initial time. 
If the term $\sin 2{\theta_0} - v_{12}(t)$ is sufficiently small during this interval of time compared to $\dot{\theta}$, 
then the adiabaticity parameter, defined in Eq.~(\ref{adpar}), is so small that the variation of the 
polarisation vector $\Delta\vec{P}$ during $\Delta t$ is negligible, as also clearly evident from the solution we have obtained
in Eqs.~(\ref{solutionPxPy})-(\ref{Pzderivative}).  In this situation it is clear that $V(t)$ does not affect the solution, and one simply recovers
the time-independent case. 

Therefore, one expects that for smaller and smaller values of $\gamma_{\rm res}$, there is a departure from the LZSM approximation and
a transition toward the time-independent solution. This has been for example studied in the case of neutrino mixing for $v_{12}(t)=0$
and an interpolating approximate {\em ansatz} solution has been given in \cite{Kuo:1988pn}. 
Our solution is derived from the density matrix equation and much more general, since it is valid also for $v_{12}(t) \neq 0$. 
Let us also notice that, though we have focused on the resonant case, the solution we have obtained
also holds in the non-resonant case. Of course both cases give the same extremely non-adiabatic limit since this, as just discussed,
simply recovers the time-independent case.

\section{Including decoherence}

Realistic quantum systems are never completely isolated from the environment but should be rather described as open systems.
In this case quantum decoherence \cite{Schlosshauer:2019ewh}, describing the effects of the interaction of a quantum system 
with the environment, needs to be taken into account.  This can be done adding a  term in the right-hand side of the von Neumann
equation that has the effect to damp the density matrix off-diagonal terms,  
obtaining the Lindblad equation \cite{Lindblad:1975ef,Gorini:1975nb,franke}
\be\label{lindblad}
{\partial \rho(t) \over \partial t} = - i \, \left[H(t),\rho(t) \right] - \Gamma(t) \,  [\sigma_z,[\sigma_z,\rho(t)]]  \,  .
\ee
In the vectorial representation this corresponds to an extended version of the Bloch equation (see Eq.~(\ref{precession})) given by
\be\label{lindblad2}
{\partial\vec{P}(t) \over \partial t} = \vec{V}(t) \times \vec{P}(t) - \Gamma(t) \,  [P_x \, \hat{x} +P_y \, \hat{y}] \,   .
\ee
The solution for $P_x$ and $P_y$ given by Eqs.~(\ref{solutionPxPy}) gets then easily extended:
\bea\label{solutionPxPybis}
P_x(t) & = &  \int_{0}^t dt' \, V_x(t') P_{z}(t') \, e^{-\Gamma(t')\, t'} \, \sin\left(\int_{t'}^t dt'' V_z(t'') \right)\,  , \\
P_y(t) & = &   -\int_{0 }^t dt' \, V_x(t') P_{z}(t') \, e^{-\Gamma(t')\, t'} \, \cos\left(\int_{t'}^t dt'' V_z(t'') \right) \,  .
\eea
Consequently, also Eq.~(\ref{Pzderivative}) and Eq.~(\ref{appearance}) get modified accordingly: 
\be\label{Pzderivative2}
{dP_z(t) \over dt}   =  -V_x(t) \, \int_{0 }^t dt' \, V_x(t') P_{z}(t') \, e^{-\Gamma(t')\, t'} \, \cos\left(\int_{t'}^t dt'' V_z(t'') \right)  \,  .
\ee 
and
\be\label{appearance2}
P_{1\ra 2}(t) =\rho_{22}(t) = {1 - P_z(t) \over 2} = {1\over 2}\, \int_{0 }^t dt' \, V_x(t')  
\int_0^{t'} dt'' \, V_x(t'') \, e^{-\Gamma(t'')\, t''} \, \cos\left(\int_{t''}^{t'} dt''' V_z(t''') \right) \,  . 
\ee
The solution is now even more general, including also the effect of decoherence, if this is not negligible.
This shows how a density matrix formalism approach allows a much more general description of 
transitions in a two-state quantum system.

\section{Conclusions}

We discussed a description of non-adiabatic transitions in two-state quantum systems within the density matrix formalism.
 After having introduced the density matrix formalism, setting the general framework, we first discussed preparatory 
 time-independent and adiabatic cases and then we derived quite a general formula for the transition probability working 
 also in the non-adiabatic case and even  beyond the usual LZSM approximation. 
 We neglected a possible azimuthal angle $\alpha(t)$, that is equivalent to have
considered only real Hamiltonians. The presence of an azimuthal angle would generate additional terms proportional to $P_y$ in the right-hand side of Eqs.~(\ref{unrotated}). The time derivative of $\alpha(t)$ would also give an additional contribution to the adiabaticity parameter. 
It will be worth to investigate an extension of our analytical procedure in this direction.  Moreover the result has been obtained
for small mixing angles, so that we could approximate $P_z \simeq 1$ in the expressions for $P_x$ and $P_y$.  However, this is the correct
approximation to study a regime where the LZSM approximation breaks down and one has a transition toward an extreme non-adiabatic limit. 
In any case it would be interesting to investigate whether it is possible to extend the expression also in this respect, relaxing the approximation $P_z \simeq 1$.   
We used, throughout the paper, solar neutrino mixing as a benchmark  playground. However, our approach, and the way we presented, 
is very general and applicable to a large variety of physical phenomena
based on non-adiabatic transitions in two-state quantum systems. The significance of our result can be understood only 
deriving results case by case. We are certainly aware that this is going to be particular important within various applications in particle mixing:
i) solar active-sterile neutrino mixing; ii) active-sterile neutrino mixing in the early universe; iii) very heavy dark matter
production from sterile-sterile neutrino mixing; iv) photon-dark photon mixing in the early universe and astrophysical environments.
It will be intriguing to explore these potential important applications in particle mixing.  
However, we envisage  that our findings could also be useful in other contexts completely different from particle mixing, such as nuclear magnetic resonance and 
quantum gates. It will be interesting to apply our approach also to these physical problems.

\vspace{-1mm}
\subsection*{Acknowledgments}

PDB and SP acknowledge financial support from the STFC Consolidated Grant ST/T000775/1.
YLZ was partially supported by National Natural Science Foundation of China (NSFC) under Grant 
Nos. 12535007, 12547104, and Zhejiang Provincial Natural Science Foundation of China under Grant No. LDQ24A050002. 
We wish to thank Amol Dighe and Ivan Martinez-Soler for stimulating discussions during WHEPP 2025 and PASCOS 2026, respectively. 

\section*{Appendix: LZSM approximation in the Hamiltonian formalism}
\appendix
\renewcommand{\thesection}{\Alph{section}}
\renewcommand{\thesubsection}{\Alph{section}.\arabic{subsection}}
\def\theequation{\Alph{section}.\arabic{equation}}
\renewcommand{\thetable}{\arabic{table}}
\renewcommand{\thefigure}{\arabic{figure}}
\setcounter{section}{1}
\setcounter{equation}{0}

In this Appendix we briefly review the derivation of the LZSM formula for the crossing probability in the usual Hamiltonian formalism.
We first review the derivation given by Zener in \cite{Zener:1932ws} and then a general derivation from semiclassical approximation \cite{pokrovskykhalatnikov,dykhne,Pokrovsky:2010zz}
that greatly simplifies the original derivation by Landau \cite{Landau:1932wdt}.\footnote{A detailed discussion on the Landau derivation \cite{Landau:1932wdt}, 
based on analytical continuation  leading to a calculation of the crossing probability 
as a solution of a path integral in the complex plane, can be found in \cite{Kuo:1988pn}.}

\subsection*{LZSM as asymptotic limit of Weber equation solution}

Let us start from Zener's derivation. Let us describe the time-evolution of the state vector in the interaction basis. 
To this extent, we can expand it in this basis writing
\be
\ket{\psi(t)} = \psi_1(t) \ket{\lambda_1} + \psi_2(t) \ket{\lambda_2} \,  ,
\ee
where $\psi_1(t) \equiv \brac{\lambda_1}\ket{\psi(t)}$ and $\psi_2(t) \equiv \brac{\lambda_2}\ket{\psi(t)}$.
Let us also define the column vector $\psi(t) = (\psi_1(t),\psi_2(t))^T$. In this way we can write 
the Schr\"{o}dinger equation in the interaction basis as:\footnote{This is fully equivalent to the von Neumann
equation for the density matrix equation in the interaction basis Eq.~(\ref{vonneumann2}).}
\be\label{SEint}
i \, {\partial \psi(t) \over \partial t} = H(t) \psi(t) \,  .
\ee
Using the decomposition of the Hamiltonian matrix Eq.~(\ref{Hdecomposition}), we know that the first term 
yields an overall phase factor, so that the solution of (\ref{SEint}) can be written in the form 
\be
\psi(t) = \widetilde{\psi}(t) \, e^{-i \int_{0}^t dt' \varphi(t')} \,  ,
\ee
where $\varphi(t) \equiv (E_1(t) +E_2(t))/2$. In this way, from Eq.~(\ref{SEint}), 
the time-evolution of $\widetilde{\psi}(t)$ is determined only by the effective Hamiltonian, 
explicitly:
\be\label{SEeff}
i \, {\partial \widetilde{\psi}(t) \over \partial t} = \Delta H(t) \, \widetilde{\psi}(t) \,  .
\ee
The effective Hamiltonian matrix $\D H(t)$ in the interaction basis is given by Eq.~(\ref{effectiveH}) or also Eq.~(\ref{effectiveH2}),
that we both rewrite here for convenience (for $\alpha =0$):
\bea\label{effectiveHbis}
\D H(t)  & = &  - {1 \over 2}\, \D E(t) \, \left(
\begin{array}{cc}
\cos 2\theta(t) & -\sin 2\theta(t) \\
-\sin 2\theta(t) & -\cos 2\theta(t) 
\end{array}
\right) \\
& = & - {1 \over 2}\, \D E_0 \left(
\begin{array}{cc}
\cos 2\theta_0 - \D v(t)& -\sin 2\theta_0 +v_{12}(t) \\
-\sin 2\theta_0 + v_{12}^\star(t)& - \cos 2\theta_0 + \D v(t)
\end{array}
\right) \, .
\eea
If $\theta(t) =0$, implying $\theta_0 = v_{12}(t) = 0$, then $\D H_{12} = 0$, and from Eq.~(\ref{SEeff}) one obtains a simple uncoupled
system of differential equations
\bea
{\partial \widetilde{\psi}_1(t) \over \partial t} & = & -i \D H_{11}(t) \widetilde{\psi}_1(t) \\
{\partial \widetilde{\psi}_2(t) \over \partial t} & = & -i \D H_{22}(t) \widetilde{\psi}_2(t)  \,   ,
\eea
with solution
\bea
\widetilde{\psi}_1(t) & = & \widetilde{\psi}_1(0)\,e^{-i\, \int_{0}^{t} dt' \D H_{11}(t')} \\
\widetilde{\psi}_2(t) & = & \widetilde{\psi}_2(0)\,e^{+i\, \int_{0}^{t} dt' \D H_{11}(t')}  \,  ,
\eea
where we used $\D H_{22}(t) = - \D H_{11}(t)$.
If the initial condition is $\psi(0) = (1 \; 0)^T$, one obtains a simple solution
\bea
\widetilde{\psi}_1(t) & = & e^{-i\, \int_{0}^{t} dt' \D H_{11}(t')} \,  ,\\
\widetilde{\psi}_2(t) & = & 0  \,  ,
\eea
implying that there would be no oscillations or conversions and, trivially, 
$P_{12}(t) = |\psi_2(t)|^2 = 0$.  If we turn on a non-vanishing
mixing angle $\theta(t)$,  Eq.~(\ref{SEeff}) now yields a system of coupled differential equations
\bea\label{csystem}
{\partial \widetilde{\psi}_1(t) \over \partial t} & = & -i \, \D H_{11}(t) \widetilde{\psi}_1(t) -i \, \D H_{12}(t) \widetilde{\psi}_2(t)\\
{\partial \widetilde{\psi}_2(t) \over \partial t} & = & +i \, \D H_{11}(t) \widetilde{\psi}_2(t)  -i \, \D H_{12}(t) \widetilde{\psi}_1(t) \,  .
\eea
These can be recast as 
\bea\label{csystem2}
\left({\partial \over \partial t} + i \, \D H_{11}(t) \right)\, \widetilde{\psi}_1(t) & = &  -i \, \D H_{12}(t) \widetilde{\psi}_2(t)\\
\left({\partial \over \partial t} - i \, \D H_{11}(t) \right)\, \widetilde{\psi}_2(t) & = &  -i \, \D H_{12}(t) \widetilde{\psi}_1(t) \,  .
\eea
From the second equation we can then write
\be
\widetilde{\psi}_1(t) = {i \over \D H_{12}(t)}\,\left({\partial \over \partial t} - i \, \D H_{11}(t) \right)\,\widetilde{\psi}_2(t) \,  .
\ee
When this is plugged into the first equation, one finds:
\be
{\partial^2 \widetilde{\psi}_2(t) \over \partial t^2}  
+ \widetilde{\psi}_2(t)\left( \D H^2_{12}(t) -i\,\D \dot{H}_{11}(t) + \Delta H^2_{11}(t) \right)   =  0 \,  . 
\ee
If we now make the assumptions that $v_{12}(t) =0$, so that 
\be
\D H_{12} = {1\over 2}\Delta E_0 \sin 2\theta_0 = {\rm const}
\ee
and that the resonant region is sufficiently small that one can Taylor expand 
\be
\D H_{11}(t) \simeq \Delta H_{11}(t_{\rm res}) + \Delta \dot{H}_{11}(t_{\rm res}) (t-t_{\rm res}) \,  ,
\ee
implying
\be
\Delta \dot{H}_{11}(t) \simeq \Delta \dot{H}_{11}(t_{\rm res}) ={1\over 2}\, \Delta E_0 \, \Delta\dot{v}(t_{\rm res}) \,   ,
\ee
the equation transforms into {\em Weber's equation} 
\be
{\partial^2 \widetilde{\psi}_2(t) \over \partial t^2}  
+ \widetilde{\psi}_2(t)\left(f^2 -i {\alpha \over 2} + {\alpha^{2}\over 4} (t - t_{\rm res})^2 \right) =  0 \,  ,
\ee
where, following Zener's notation, we defined $f = \D H_{12}$ and  $\alpha = 2 \Delta \dot{H}_{11}(t_{\rm res})$ (not to be
confused with the azimuthal angle introduced in the body text).
Notice that the adiabaticity parameter at the resonance (see Eq.~(\ref{resadpar})) can be expressed 
in terms of $f$ and $\alpha$ as $\gamma_{\rm res} = 4\, f^2/|\alpha|$. Finally, with a change of independent variable 
\be\label{t2z}
t \ra z = \alpha^{1/2}\,e^{-i{\pi\over 4}}\,\Delta t \,  ,
\ee
where we defined $\Delta t \equiv t-t_{\rm res}$,  Weber's equation reduces to its standard form
\be
{\partial^2 \widetilde{\psi}_2(z) \over \partial z^2}  
+ \widetilde{\psi}_2(z)\left( \nu + {1\over 2} - {z^2 \over 4} \right) =  0 \,  ,
\ee
with $\nu \equiv i\,f^2/\alpha$.  We are interested in the particular solution
that satisfies the boundary conditions at $t=0$, corresponding to
take the limit $\Delta t \ra -\infty$ within the linear approximation. Solutions are given by parabolic cylinders
functions $D_{\nu}(z)$, also called Weber functions.  There are two independent solutions: $D_{\nu}(z)$ and $D_{-\nu-1}(i z)$.
The Weber function $D_{-\nu-1}(iz)$ vanishes for $|z| \ra \infty$ along the Stokes lines in the complex plane
corresponding to ${\rm Arg}(z) = -3\pi/4$ and  ${\rm Arg}(z) = -\pi/4$. From Eq.~(\ref{t2z}), one can see that
the limit $\Delta t \ra -\infty$ corresponds to take $|z|\ra \infty$ and for $\alpha > 0$ one has ${\rm Arg}(-z) = -\pi/4$,
while for $\alpha < 0$ one has ${\rm Arg}(z) = -3\pi/4$. This implies that  the solution is of the form
\bea
\widetilde{\psi}_2(z) & = & A_{+} \,  D_{-\nu-1}(-iz) \,  \hspace{5mm} \mbox{\rm for} \; \alpha > 0 \,  ,  \\
\widetilde{\psi}_2(z) & = & A_{-} \,  D_{-\nu-1}(+iz) \,  \hspace{5mm} \mbox{\rm for} \; \alpha < 0 \,  . 
\eea
These solutions satisfy the boundary condition $|\widetilde{\psi}_2(\Delta t \ra -\infty)| = 0$.
The constant $A_{\pm}$ can be determined by imposing the second boundary condition  $|\widetilde{\psi}_1(\Delta t \ra -\infty)|^2 = 1$,
finding
\be
|A_-| = |A_+| = {1\over 2}\,\gamma_{\rm res}^{1\over 2} \, e^{-{\pi \gamma_{\rm res}\over 16}} \,  .
\ee
Taking then the limit of the Weber function for $\Delta t \ra \infty$, corresponding this time to take
$|z|\ra \infty$ and ${\rm Arg}(-z) = +\pi/4$ (${\rm Arg}(z) = +3\pi/4$) for $\a > 0$ ($\alpha <0$), one finally finds the LZSM approximation
\be\label{LZSMformula}
|\widetilde{\psi}_2(\Delta t \ra \infty)|^2 = 1 - e^{-{\pi \over 2}\gamma_{\rm res}} \,  .
\ee

\subsection*{LZSM from semiclassical approximation}

Let us now show that this result can also be derived in an easier way without having to know the Weber functions
following a procedure that generalises the original one by Landau. Let us start from Eq.~(\ref{csystem2}).
This time let us replace instead $\widetilde{\psi}_2(t)$ from the first of the two equations, writing
\be
\widetilde{\psi}_2(t) = {i \over \D H_{12}(t)}\,\left({\partial \over \partial t} + i \, \D H_{11}(t) \right)\,\widetilde{\psi}_1(t) \,  
\ee
and plugging  in the second one. We obtain the following second order differential equation for $\widetilde{\psi}_1(t)$:
\be
{\partial^2 \widetilde{\psi}_1(t) \over \partial t^2}  
+ \left( i\,\D \dot{H}_{11}(t) + \Delta H^2_{11}(t) + \D H^2_{12}(t)  \right) \, \widetilde{\psi}_1(t)  =  0 \,  . 
\ee
Making the same assumptions as before, and in particular that all the dynamics takes place in a narrow region
around the resonance and defining this time $\tau = \sqrt{2\dot{\D H}_{11}(t_{\rm res})}(t - t_{\rm res})$
and considering that $\gamma_{\rm res} = 2 \D H^2_{12}/\dot{\D H}_{11}(t_{\rm res})$, we arrive again to
a parabolic cylinder equation (not in standard form):
\be\label{parabolic}
{\partial^2 \widetilde{\psi}_1(\tau) \over \partial \tau^2}  
+ \left( {i \over 2} + {\tau^2\over 4} + {\gamma_{\rm res} \over 4}  \right)  \widetilde{\psi}_1(\tau)   =  0 \,  . 
\ee
This time, instead of using Weber functions and their properties, we use the much more general semiclassical approximation,
writing the solution in the form $\widetilde{\psi}_1(\tau) = e^{i {S(\tau)}}$. In this way the Eq.~(\ref{parabolic}) becomes:
\be\label{semiclassical}
\left({dS\over d\tau}\right)^2 - i {d^2 S\over d\tau^2} = {\tau^2 \over 4} + {i \over 2} + {\gamma_{\rm res} \over 4} \,  .
\ee
Since we are interesting in the limit $|\tau|\ra \infty$ for ${\rm Arg}(\tau)=0$, we can take an expansion in $1/\tau$.
The leading term $S_0$ is easily found:
\be
{dS_0\over d\tau} = -{\tau \over 2} \Rightarrow S_0 = - {\tau^2 \over 4} \,  .
\ee 
Let us now consider the next-to-leading term $S_1$, writing $S \simeq S_0 + S_1$.  The second derivative can be evaluated 
at the leading order, so that now Eq.~(\ref{semiclassical}) gives
\be
{dS_0 \over d\tau}{dS_1 \over d\tau} = {\gamma_{\rm res}\over 8} \Rightarrow S_1(\tau) = - {\gamma_{\rm res}\over 4}\ln\tau + C \,  .
\ee
We can also write explicitly:
\be
S_1(\tau) = - {\gamma_{\rm res} \over 4} \ln |\tau| - i {\gamma_{\rm res} \over 4}{\rm Arg}(\tau) + C .
\ee
The integration constant $C$ can be fixed imposing the initial condition $|\widetilde{\psi}_1(0)| = 1$, implying 
that $S_1$ needs to be real for $\tau \ra -\infty$, corresponding to ${\rm Arg}(\tau) = \pi$. This yields $C = i\pi \gamma_{\rm res}/4$.
Taking finally the limit  for $|\tau| \ra \infty$ and ${\rm Arg}(\tau) = 0$,  one finds ${\rm Im}[S_1(\tau)] = \pi \gamma_{\rm res}/4$,
yielding again  the LZSM approximation Eq.~(\ref{LZSMformula}).

\end{document}